\documentclass[12pt,reqno]{article}

\usepackage{amsfonts}
\usepackage{amsmath}
\usepackage{amssymb}
\usepackage{amsthm}
\usepackage{setspace}
\usepackage{fullpage}
\usepackage{graphicx}
\usepackage{url}
\usepackage{braket}
\usepackage{mathtools}
\usepackage[all]{xy}
\usepackage{caption}

\usepackage{color}
\usepackage{bm}

\usepackage[numbers]{natbib}

\usepackage{verbatim}
\usepackage{graphicx}

\usepackage{mathrsfs}

\theoremstyle{remark}

\theoremstyle{definition}

\newcommand{\dd}{\mathrm{d}}
\newcommand{\ee}{\mathrm{e}}
\newcommand{\pd}{\partial}
\newcommand{\Mpl}{M_{\rm Pl}}

\newcommand{\llangle}{\left\langle}
\newcommand{\rrangle}{\right\rangle}

\definecolor{darkgreen}{RGB}{50,150,0}

\begin{document}

\begin{titlepage}
\hfill \\
\vspace*{15mm}
\begin{center}
{\large \bf Topological Gravity as the Early Phase of Our Universe}

\vspace*{15mm}

{Prateek Agrawal$^{1}$, Sergei Gukov$^{2}$, Georges Obied$^{3}$ and Cumrun Vafa$^{3}$}
\vspace*{8mm}
\scriptsize{\\
$^1$Rudolf Peierls Centre for Theoretical Physics, Clarendon Laboratory, Parks Road, Oxford OX1 3PU, UK\\
$^2$Walter Burke Institute for Theoretical Physics, California Institute of Technology, Pasadena, CA 91125, USA\\
$^3$Jefferson Physical Laboratory, Harvard University, Cambridge, MA 02138, USA\\}

\vspace*{0.7cm}

\end{center}
\begin{abstract}
Motivated by string dualities we propose topological gravity as the early phase of our universe.  The topological nature of this phase naturally leads to the explanation of many of the puzzles of early universe cosmology.  A concrete realization of this scenario using Witten's four dimensional topological gravity is considered.  This model leads to the power spectrum of CMB fluctuations which is controlled by the conformal anomaly coefficients $a,c$.  In particular the strength of the fluctuation is controlled by $1/a$ and its tilt by $c g^2$ where $g$ is the coupling constant of topological gravity.  The positivity of $c$, a consequence of unitarity, leads automatically to an IR tilt for the power spectrum.   In contrast with standard inflationary models, this scenario predicts $\mathcal{O}(1)$ non-Gaussianities for four- and higher-point correlators and the absence of tensor modes in the CMB fluctuations.

\end{abstract}

\end{titlepage}

\tableofcontents

\section{Introduction}
Novel ideas coming from string theory, and in particular stringy ingredients such as branes and dualities have dramatically impacted our understanding of many areas of theoretical physics.  The same, unfortunately, cannot be said about the impact of string theory on our understanding of early universe cosmology.   Most of the effort in this direction has been to try to embed the inflationary scenario within string theory, which at best involves constraining the parameters of the model. However, these models are generally difficult, if not impossible, to construct due to various Swampland principles~\cite{Ooguri:2006in,Rudelius:2015xta,Obied:2018sgi,Bedroya:2019snp}, and thus it is natural to look for alternative stringy realizations of the early universe.
Even though there has been other efforts in applying more stringy ingredients to cosmology, there has not yet emerged a fully satisfactory stringy scenario for the early universe.  The aim of this paper is to propose a new scenario for the early universe inspired by what we have learned from string theory and in particular its dualities.

The basic idea we advance, motivated from string dualities, is that the early phase of our universe is described by a dual theory, where the matter making up the universe at early times is different from the current stage of the universe.   This idea is exemplified in string gas cosmology~\cite{Brandenberger:1988aj} where the matter we are made of (momentum modes in our space) becomes very heavy when the radius of the universe becomes stringy $R\ll l_s$ and is replaced by lighter winding string modes.  Moreover our current description of space becomes an ineffective description of the universe in the early phase, and is replaced by a dual space with radius $R'=l_s^2/R$.  In this paper we will not assume any specific form for the dual theory.  Instead, the new idea we propose here is that the early phase of any such universe as viewed from the perspective of our current universe is a topological theory where local metric fluctuations are physically absent. In particular, unlike what has been assumed thus far, we assume that Einstein's theory of gravity is replaced by a topological version in this early phase of the universe.

That a topological phase of gravity should emerge at high energies in string theory has been hinted at from many perspectives (see e.g.~\cite{Gross:1988ue,Witten:1988xi}), and given the fact that the early universe has high temperature it is not a radical proposal to assume that the early phase of our universe is described by a topological theory.  Emergence of a topological theory in the early phase of our universe naturally leads to an automatic resolution to many of the puzzles of early cosmology, including homogeneity, isotropy and scale invariance.  To illustrate more concretely how such a scenario would work, we consider a model of 4d topological gravity proposed by Witten~\cite{Witten:1988xi}.  In particular we point out that anomalous contributions to scale invariance would lead to a deviation from a scale invariant fluctuation spectrum with the tilt being towards the IR forced by unitarity of the theory.  Conformal anomaly coefficients $a$ and $c$ play a key role here in the structure of the CMB fluctuations: the magnitude of the fluctuations is controlled by $\sim 1/a$ and the tilt by $\sim cg^2$, with the positivity of $c$~\cite{Cappelli:1990yc,Duff:1993wm} (which follows from unitarity) being directly related to the IR tilt.  Tensor mode fluctuations, unlike many other models of the early universe, are absent in this scenario due to the topological nature of gravity.  Moreover we find that there are order unity non-Gaussianities in the fluctuations for four-point and higher correlators, larger than the corresponding inflationary correlators where non-Gaussianities are suppressed by slow-roll parameters.

The organization of this paper is as follows.  In section~\ref{sec:dualities} we review lessons learned from string duality in anticipation for applications to early universe cosmology.  In section~\ref{sec:puzzles} we review the basic puzzles of the early universe and the fact that many of them look like vanishing properties.  In section~\ref{sec:proposal} we propose the emergence of a topological phase as an early phase of the universe which would naturally explain such vanishing properties.  In section~\ref{sec:topgrav} we review Witten's 4d topological gravity, with emphasis on aspects relevant to our application and estimate the structure of the CMB fluctuations which they lead to.  In section~\ref{sec:compar} we contrast our model with predictions of inflationary models. In section~\ref{sec:future} we conclude with some comments about directions for future research.

\section{Dualities and early universe}
\label{sec:dualities}

One of the main lessons learned from string theory in the past couple of decades about the nature of quantum gravity is the prevalence of dualities.  Duality symmetries imply that, when we vary the parameters specifying the configurations of a given system and go to extreme corners we get totally different descriptions of the physics.  This is accomplished by the emergence of light modes in the extreme regions of parameter space, which provide a new description of the system, i.e., a dual description.  In other words, the lesson learned is that {\it there is no effective theory which is valid in all regions of a physical system as we vary the parameters of the theory}.  This breakdown of effective theory is a feature of all theories we know in the context of string theory. This is captured by the distance conjecture and is one of the Swampland principles~\cite{Ooguri:2006in}.   This in particular implies that as the dimensionful parameters of the system change by $\mathcal{O}(1)$ in Planck units, an exponentially light tower of states emerge.  In particular, {\it an ``all encompassing'' effective theory which is valid in all regimes of parameter space --- even though this may naively sound to be a better situation to be in --- is believed not to be a consistent quantum gravitational theory and belongs to the Swampland.}
Let us revisit the situation with early cosmology in light of this general lesson learned from string theory.

If we consider the Friedmann-Lema\^itre-Robertson-Walker (FLRW) cosmology which provides a good description of cosmology except possibly for very early times, we see that the temperature of the universe is decreasing as time evolves.  Moreover as we extrapolate back in time the temperature of the universe is expected to increase even beyond the Planck temperature.  Of course we do not expect FLRW to be valid in this regime.  How do we deal with this situation?  One possible solution, as in the inflationary scenario~\cite{Guth:1980zm,Linde:1981mu,Albrecht:1982wi}, or bounce scenarios~\cite{Steinhardt:2001st,Khoury:2001wf,Donagi:2001fs} is to include additional terms or fields in the action that become important when we go back to early times.  In all these scenarios, the additional fields or terms one considers can still be described by the same effective Lagrangian, including early and late times.  However, this is not what string dualities would have led us to expect:  we would have expected that in early times we have a totally different dual description of the physics which could not have been captured by the addition of a few fields or terms to the Lagrangian. In particular, the limit of infinite temperature is an infinite distance limit from the viewpoint of the Swampland distance conjecture (DC)~\cite{Ooguri:2006in}. One can see this by recalling that a finite temperature $T = \beta^{-1}$ is implemented by a Euclidean time circle with radius $R_t = \beta/2\pi$. The distance relevant for the DC is then $\sim |\log \beta|$ so we expect a tower of states of mass:
\begin{align*}
  m \sim \exp(-a |\log \beta|) \sim T^{-a}.
\end{align*}
The strong version of the DC then implies the existence of a dual description as $T\rightarrow \infty$. It was with this motivation in mind that the model of string gas cosmology was proposed for the early universe~\cite{Brandenberger:1988aj}.  Even though we will not be using this specific model in this paper, it sets the stage for some of the motivations for what we will be using. We will now turn to a review of string gas cosmology.

\subsection{String Gas Cosmology}\label{sec:stringGas}

Consider the universe placed in a periodic 3-dimensional box of length $R$. For concreteness let us consider the heterotic string.  The other dimensions are viewed as compact and stabilized so we will not worry about their dynamics.  In an FLRW cosmology,
as we go backwards in time $R$ gets smaller.
We expect that when $R\lesssim l_s$ a new description should take over, where $l_s$ is the string scale. Indeed, this is one of the basic examples of string duality, known as T-duality:  the theory is dual to strings propagating in a box of size $l_s^2/R$, and so in this dual description we have a universe which is expanding as $R$ decreases.  Note that this is not a `bounce' scenario~\cite{Steinhardt:2001st,Khoury:2001wf,Donagi:2001fs}, as in terms of the original variables the box is still contracting.  But in a dual description it is better described by an expanding universe.

This behavior is surprising from the point of view of a theory of point particles.   In the first description, string momentum modes are the light degrees of freedom ($E\sim n/R$) used to define position variables
$$|x\rangle = \sum_p e^{i p x}|p\rangle.$$
However, in the dual description, it is the string winding modes (absent in a theory of point particles) that are the appropriate light degrees of freedom to define position variables
 $$|\tilde{x}\rangle = \sum_w e^{i w  \tilde{x}} |w\rangle,$$
that is because the winding modes have energy $E\sim R/l_s^2\ll 1/R$.  So, as we go back in time where $R\ll l_s$  the momentum modes, which constitute the degrees of freedom our universe is made of, get converted into the lighter winding modes. This is consistent with the temperature dropping down as we go back in time, even though we would have naively expected, based on FLRW cosmology for it to continue to increase. From this viewpoint, our universe emerged from the crunched limit of a dual universe, whose local dynamics captured in terms of spatial variables ${\tilde x}$ cannot be expected to be described by local physics in our universe, and the physics of that earlier time cannot be described using only the fields we are familiar with in our universe.  It is clear in this context that we do not have a complete single patch description of all physics for all times.

Even though this model has been used with additional assumptions~\cite{Tseytlin:1991xk,Brandenberger:2006vv,Nayeri:2005ck,Nayeri:2006uy} to lead to quantitative predictions, these assumptions cannot be fully justified (see e.g. \cite{Kaloper:2006xw}).
Indeed the main issue in extracting predictions from this model such as gravitational fluctuations is that we cannot go back in time and expect the notion of an effective Einstein gravity coupled to matter, valid in the current phase of our universe, to be valid at very early times (see however~\cite{Bernardo:2019pnq,Bernardo:2020nol} for an attempt in this direction).   It is more natural to ask whether there is a more general framework where the lessons of string dualities can be used, without any further assumptions tied to a given model.   It is the aim of this paper to propose such a framework which captures some of the basic features inspired from this model and more generally from string dualities.

\section{General puzzles of the early universe and dual descriptions}
\label{sec:puzzles}
We begin by recalling the puzzles of modern cosmology (see~\cite{Baumann:2009ds} for a pedagogical introduction) and noting that they all can be rephrased as vanishing statements when certain operators act on the state of our universe. We discuss homogeneity, isotropy, flatness and scale-invariance and regard perturbations (and their deviation from scale-invariance) as small effects that can arise by slight deformations of models that exactly produce the above properties.

\subsection{Homogeneity and isotropy}
This is also called the horizon problem which is the puzzle arising from the observation that the universe is homogeneous across patches that have been causally disconnected since the onset of Big Bang cosmology. Another way to phrase the homogeneity problem is to ask: why is the state of the universe translationally invariant?

In quantum mechanics, a state is translationally invariant if it is annihilated by the momentum operator\footnote{Here we use the language of flat space but there would be analogous operators on curved spaces such as the angular momentum operator on a sphere.}:
\begin{align}
  \label{eq:translationalInv}
  \hat{\bf p}|\psi \rangle = 0.
\end{align}
In addition, if the momentum operator commutes with the Hamiltonian
\begin{align}
  \label{eq:commutator}
  [\hat{\bf p}, \hat{H}] = 0,
\end{align}
then time evolutions of the state $|\psi\rangle$ are also translationally invariant as one can readily check. Here, it is important to note that condition~(\ref{eq:translationalInv}) is a property of the state under consideration, whereas the commutator~(\ref{eq:commutator}) is a property of the theory and therefore holds more generally. This observation guarantees that if one were able to prepare a universe in a translationally invariant initial state (i.e. one that is annihilated by the momentum operator) then that universe would have no horizon problem.

Two comments are in order here. First, the Hamiltonian above is taken to govern the evolution of matter on a background spacetime and is not the Hamiltonian of gravity which would vanish by time reparameterization invariance. This is consistent with the discussion of energy densities in cosmology. Second, the vanishing of the commutator is only needed to hold in a short time interval. In an expanding universe, momentum and energy density are expected to be conserved only on time scales short compared to the Hubble expansion rate and it is in this sense that we expect the vanishing of the commutator above which is intended as statement of instantaneous energy conservation.

\subsection{Flatness}
The flatness problem is a fine-tuning problem that arises from the observation that the universe must have started with an energy density very close to the critical value in order to produce the spatially flat universe we observe today. More precisely, the curvature density parameter $\Omega_k \equiv -k(aH)^{-2}$ has been observationally constrained to be less than $\sim 10^{-3}$~\cite{Aghanim:2018eyx}. This is fine-tuned because $\Omega_k = 0$ is an unstable fixed point of the FLRW cosmology and the observed small number today implies that $|\Omega_k | \lesssim 10^{-16}$ at BBN for example or $|\Omega_k| \lesssim 10^{-55}$ at the GUT scale. Similar to the case of homogeneity above, flatness is the statement that local spatial curvature vanishes. This can be understood as the vanishing commutation of the generators of translation.

\subsection{Scale invariance}

The homogeneous early universe contained small perturbations that form the seed for structure formation. The power spectrum of these perturbations is the variance of the perturbation amplitude as a function of its comoving wavenumber. A scale invariant perturbation spectrum is one where the two-point function of the gravitational potential\footnote{We only consider the two-point function on superhorizon scales in this case. This is a constant multiple of the gauge-invariant curvature perturbation.} does not depend on comoving coordinate $x$.  Let us denote the scalar part of the metric fluctuation by $\hat\Phi=\delta g/g$.
Considering a system in a state $|\psi\rangle$, independence of the two-point fluctuations of the metric on $x$ can be rephrased as the vanishing statement:
\begin{align}
  \langle \psi| [D, \hat{\Phi}(0) \hat{\Phi}(x)] |\psi \rangle = 0.
\end{align}
where $D$ is the dilation operator.  This is an approximate symmetry of the observed power spectrum in our universe.  More specifically it has been found that the power spectrum has a small infrared tilt.  Namely
\begin{align}
k^3\langle \psi| \hat{\Phi}(\vec{k}) \hat{\Phi}(-\vec{k}) |\psi \rangle' \sim k^{-0.03}
\end{align}
where the prime denotes that we have removed an overall momentum-conserving Dirac delta function.

\section{Proposal for an early universe scenario:  the topological phase}
\label{sec:proposal}
 In this section we first explain how the string gas cosmology scenario fits with resolving the early universe puzzles and use this to propose a general scenario for the early universe and in particular the emergence of a topological phase of gravity.
 \subsection{String duality and the topological phase}
The requirements of homogeneity and scale invariance are not exotic from the point of view of string dualities. We illustrate this by taking T-duality, which is a key ingredient in string gas cosmology, as an example and focusing on translation invariance. As previously mentioned, states in the Hilbert space of string theory on a torus are described by momentum and winding quantum numbers (in addition to others that are not essential for this discussion). Their contribution to the mass is proportional to $R$ and $R^{-1}$ respectively. Hence, when the torus is at small radius $R \ll l_s$, one would expect a state built only from winding modes since momentum modes are heavy in comparison. A natural initial state is then:
\begin{align}
  |\psi_i\rangle = \sum_i c_i |w_i, p = 0\rangle
\end{align}
where $w$ and $p$ schematically denote winding and momentum quantum numbers and $c_i$ are arbitrary coefficients. This state clearly satisfies the condition in Eq.~\eqref{eq:translationalInv} and would thus automatically lead to a homogeneous universe.

T-duality is used here as an illustration but the main principle holds more generally as we now argue. Suppose we have fields $\hat{\mathcal{O}}$ whose non-zero eigenmodes are light in one phase (let's call it phase II) and heavy in another phase (phase I). Starting with a cosmology in phase I, the observables of phase II modes will be energetically disfavored and thus remain unpopulated. The natural state is then one that is annihilated by the operator $\hat{\mathcal{O}}$ as only the zero mode can be occupied. Time evolution can then reshuffle the mass spectrum exchanging heavy and light states and necessitate a transition to duality frame II. As the evolution of the universe proceeds, modes that were initially populated become heavy in frame II and decay into the lighter non-zero eigenmodes of $\hat{\mathcal{O}}$. To low-energy observers who may not know about the existence of dualities, it is light modes in frame II that are useful for describing the universe. The presence of non-trivial correlations between these light modes would be puzzling but not so once they realize that all light modes emerged from decay of ``invisible'' modes (the light modes in frame I) which are momentum zero modes as far as frame II is concerned.  In other words they look like global/topological modes from the perspective of frame II.  In particular {\it the resolution of homogeneity and horizon problem will naturally be related to the fact that the local physics description of phase II breaks down in phase I}, where new degrees of freedom describe the physics, and during the conversion process from phase I to phase II translation symmetry is a good symmetry leading to a homogeneous universe.  So the answer as to which physical process led to homogeneity in our universe, is that the physics is non-local from the perspective of frame II, but local from the perspective of the correct duality frame for early universe, which is frame I.  Thus the basic summary of this scenario is that {\it from the perspective of frame II, the matter in phase I cannot be well described and the spatial features of the theory from the perspective of frame II are captured by a topological theory.}

It is possible to generalize the above argument to also imply isotropy of the universe. The fact that the degrees of freedom of phase I are difficult to excite in phase II means that phase I would not have any local excitations when viewed from phase II. This amounts to saying that it is impossible to pinpoint specific locations on the spacelike slice between the two phases. It is then clear that observables also cannot pick out directions on the spacelike slice hence implying the isotropy of our universe.

Thus we see that the stringy perspective on the cosmological puzzles is that the difficulty in reconciling observations related to early universe with the usual Einstein theory coupled to matter is to be expected:  Puzzles arise if one insists on finding a {\it single} effective description valid in all regimes.  In other words, {\it the emergence of cosmological puzzles from any given perspective is a prediction of the stringy perspective that any effective theory with a finite number of degrees of freedom breaks down in extreme regions of parameter space.}

 \subsection{Basic outline of the scenario}
\begin{figure}
\centering
\includegraphics[width=\textwidth]{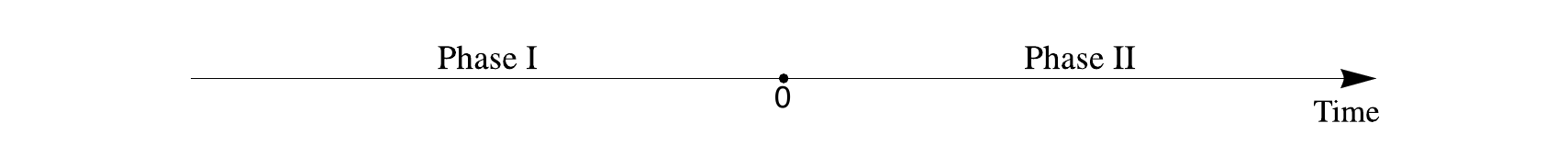}
\includegraphics[width=\textwidth]{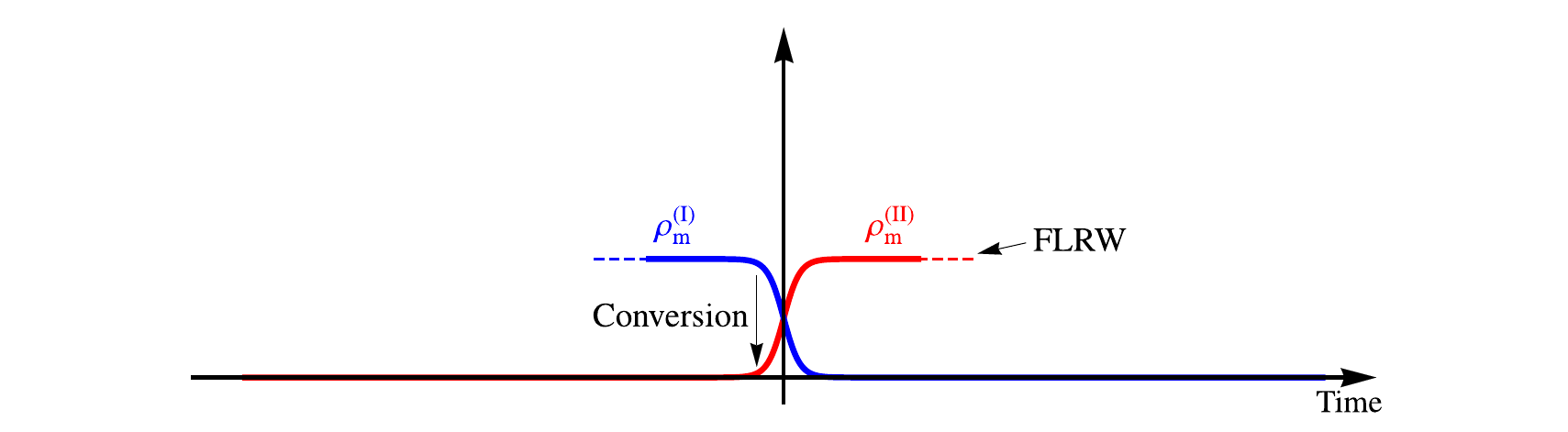}
\caption{Transition from phase I to II in our universe proceeds by the conversion of matter made up from the degrees of freedom of a dual frame (blue) to those of our frame (red). As the universe expands, heavy and light degrees of freedom are shuffled causing the conversion. }
\label{fig:phaseIandII}
\end{figure}

Motivated by string dualities, exemplified in the context of the string gas cosmology, and the puzzles associated with early universe, we propose the following scenario for early times:  We propose that as we go back in time and our universe contracts, we will reach a point where the degrees of freedom that our universe is made of disappear and get replaced by other light degrees of freedom.  Going forward in time, this means that we will be assuming that our universe emerged from a phase, which we shall call phase I, where the degrees of freedom of that universe became heavy (``dinosaurs'') and got converted to the degrees of freedom we are made of as they become lighter, in phase II (see Figure~\ref{fig:phaseIandII}).  Let us call this time $t=0$.  In other words, we will be assuming that we have two phases, depicted by $t<0$ and $t>0$.  Our universe gets populated with degrees of freedom for $t>0$ whereas $t<0$ is a phase where another universe made of totally different degrees of freedom existed. Moreover, we do not expect a single effective theory of gravity to describe both phases.  However, in order to be able to make predictions about the structure of matter in our universe, i.e. phase II,  we need to make some plausible assumptions about how some aspects of phase I looks like from the vantage point of duality frame II.

Note that the notion of time is common to both phases of the universe.  And so this leads to a notion of energy common to both phases.
The most natural assumption in the conversion of matter from the previous phase to our phase is that the matter gets converted randomly to all possible degrees of freedom of phase II, compatible with available energy.  In other words, the matter that we get in phase II for $t>0 $ will be well described by a thermal distribution of matter on top of which we have superimposed weak long range correlations that originate from modes that are non-local in phase II. This is indeed a very good approximation to the initial conditions we need for FLRW cosmology.
The horizon problem is solved simply because the locality which is relevant in our universe in phase II is not natural in phase I. In other words, the physics and light modes of phase I, just as in the winding modes of the string gas cosmology, are {\it non-local} as viewed from the perspective of phase II and fluctuations visible in phase II are not part of the degrees of freedom of phase I.

\subsubsection{Phase I as viewed from frame II perspective: A topological theory}

\begin{figure}
\centering
\includegraphics[width=\textwidth]{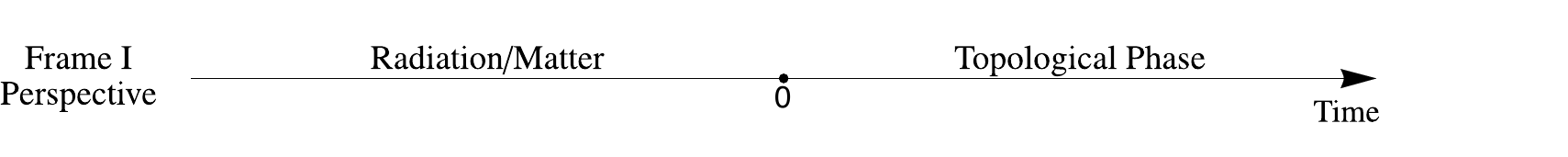}
\includegraphics[width=\textwidth]{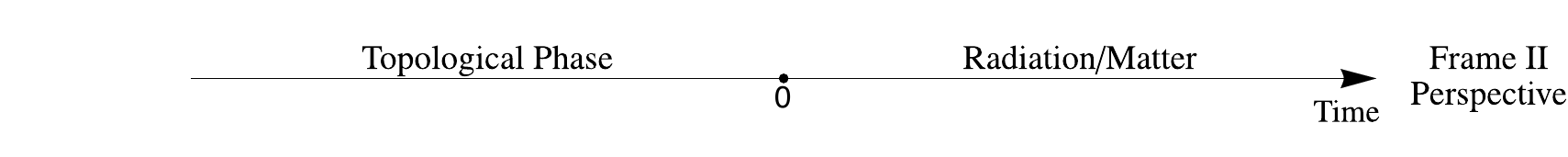}
\caption{Since the degrees of freedom making up phase I are absent in a low energy description of phase II, the former appears topological from the point of view of the latter. This relation is also true with the roles of phase I and II interchanged.}
\label{fig:perspective}
\end{figure}

Even though phases I and II may be similar, since our description of phase II uses different degrees of freedom we would need to ask how phase I looks from the perspective of frame II?  Indeed in phase I to a good approximation there should not be any position dependent observables because all the position dependent excitations are turned off.  Let us assume the state in phase I is given by a state $| I\rangle$. We would expect $n$-point correlations of suitable observables in this state
$$\langle I | {\cal O}^{i_1}(x_1)\ldots {\cal O}^{i_n}(x_n)|I \rangle =A^{i_1,\ldots,i_n}$$
to be position independent $\partial _j A^{i_1,\ldots,i_n}=0$.  This is the hallmark of a topological quantum field theory.  We are thus led to view phase I as a topological phase as viewed from the perspective of frame II. It is curious to note that the reverse is also true:  namely, phase II can be viewed from the perspective of frame I as a topological theory. This is illustrated in Figure~\ref{fig:perspective}.

The existence of a topological phase in the early universe is the main proposal we make in this paper and we explore consequences of this assumption.  As already described, the horizon problem is automatically solved in this framework. In order to explore other aspects of the early universe, such as flatness and near scale-invariance, we need to have a more precise description of phase I.
Here we are on a limb, as we do not know which topological system we have.  We do know that we want gravity to be a part of it. In other words, we need a topological gravity theory in 4 dimensions.  We will mainly consider a simple model for 4d topological gravity proposed a long time ago by Witten~\cite{Witten:1988xi} whose 2d cousin has been very well studied.  It would also be interesting to explore other such models of 4d topological gravity in view of their potential application to early universe cosmology.

In the next section we review Witten's 4d topological gravity. After that we argue how flatness can be a natural consequence of that model. Moreover, the theory being topological starts with a built in scale invariance, which gets broken.  We will also explore the broken nature of this scale invariance, and how the 4d topological gravity of Witten enjoys an approximate scale invariance. In this way we address the additional two puzzles of the early universe cosmology.

\section{4d topological gravity}
\label{sec:topgrav}
An important aspect of topological theories is the absence of local excitations so that observables are a measure of global features. This also implies that the propagation of signals is a meaningless concept in topological theories thereby automatically giving a solution to the horizon problem as we mentioned earlier. This property is achieved in the theory\footnote{Here we mainly focus on the so-called Witten-type (also known as ``cohomological'') theories, but one can also consider the generalization to Schwarz-type theories. For a review see~\cite{Birmingham:1991ty}.} by the presence of a Grassmann odd BRST charge operator $Q$ which is nilpotent (up to, perhaps, a gauge transformation). The action of the BRST charge on fields $\Phi$ is given by $\delta\Phi = i \epsilon [ Q,\Phi \}$ and pairs together bosonic and fermionic fields in a way similar to the pairing in supersymmetric theories. Physical states in the Hilbert space are $Q$-cohomology classes: these are states that are $Q$-closed (i.e. $|\psi\rangle$ satisfying $Q|\psi\rangle=0$) modulo $Q$-exact (i.e. $|\psi\rangle$ such that $|\psi\rangle = Q|\chi\rangle$ for some $|\chi\rangle$). This latter requirement implies that the fermionic partners of bosonic fields are in fact ghosts so that all degrees of freedom cancel in the BRST sense.

If we assume that the vacuum is $Q$-invariant, then it is easy to see that $Q$-exact operators have a vanishing expectation value $\langle [ Q,\mathcal{O} \}\rangle = 0$. In topological field theories, the energy-momentum tensor (given by the variation of the action with respect to the metric) is $Q$-exact, i.e. $T_{\alpha\beta} = \{Q, \lambda_{\alpha\beta}\}$ for some $\lambda$. This implies that the partition function is invariant under metric variations:
\begin{align*}
    \delta Z &= \int \mathcal{D}\Phi\, \ee^{-S}\left(-\delta S\right)
    = - \int \mathcal{D}\Phi\, \ee^{-S}\{Q,\int \sqrt{g}\delta g^{\alpha\beta} \lambda_{\alpha\beta}\}\\
    &= -\langle \{Q,\int \sqrt{g}\delta g^{\alpha\beta} \lambda_{\alpha\beta}\} \rangle = 0
\end{align*}
provided the integration measure is BRST invariant.

In topological gravity, the partition function $Z$ includes an integral over the metric degrees of freedom $g$ (discarding gauge redundancies as usual). It is thus automatically independent of any background metric. However, there is still a question of whether the theory contains local excitations. Witten's theory~\cite{Witten:1988xi}, can be obtained by applying the Batalin-Vilkovisky (BV) formalism~\cite{Batalin:1981jr,Batalin:1984jr,Batalin:1985qj} to the topological action $W\wedge W$ where $W$ is the Weyl tensor~\cite{Brooks:1988jm}. Since the theory we start with is locally trivial, the final result is an action containing bosonic fields and ghosts so that the local dynamics is absent. If one ignores the ghosts, then the dynamics of gravity would be governed by a self-dual Weyl action
\begin{equation}
S_g = \int \dd^4x \sqrt{g}\, \frac{1}{2}(W + \star W)^2
\label{WWaction}
\end{equation}
where $\star$ is the Hodge dual. This action is scale-invariant classically but would generally have a conformal anomaly at the quantum level (see below).
In addition, conformal symmetry is broken in Witten's topological gravity by a vev of a scalar field (denoted $\Phi$ in \cite{Witten:1988xi}) that is required for the action to be non-degenerate. Despite the fact that the usual Weyl tensor square gravity has ghosts and is non-unitary, this topological theory is unitary~\cite{Witten:1988xi} as the non-unitary correlations are not allowed observables of the topological theory. We also note that the Einstein-Hilbert term $\mathcal{R}$ is not generated in Witten's topological gravity because it is forbidden by the BRST symmetry.
In order to see this, we need to review in more detail the field content and the BRST transformations.

\begin{table}[ht]
	\begin{centering}
		\begin{tabular}{|c||c|c|}
			\hline
			$\phantom{\int^{\int^\int}} \text{field} \phantom{\int_{\int}}$~ & ghost number & ~~$[ Q , \text{field} \}$~~ \tabularnewline
			\hline
			\hline
			$\phantom{\int^{\int^\int}} C_{A\dot A} \phantom{\int_{\int}}$ & $2$ & $\psi_{AB,\dot A \dot B} C^{B \dot B}$
			\tabularnewline
			\hline
			$\phantom{\int^{\int^\int}} \psi_{AB, \dot A \dot B} \phantom{\int_{\int}}$ & $1$ & $- \frac{i}{4} \big( e^{\alpha}_{A \dot A} D_{\alpha} C_{B \dot B} + e^{\alpha}_{B \dot A} D_{\alpha} C_{A \dot B} + e^{\alpha}_{A \dot B} D_{\alpha} C_{B \dot A} + e^{\alpha}_{B \dot B} D_{\alpha} C_{A \dot A} \big)$
			\tabularnewline
			\hline
			$\phantom{\int^{\int^\int}} e_{\alpha A \dot A} \phantom{\int_{\int}}$ & $0$ & $e_{\alpha}^{B \dot B} \psi_{AB,\dot A \dot B}$
			\tabularnewline
			\hline
			$\phantom{\int^{\int^\int}} W_{ABCD} \phantom{\int_{\int}}$ & $0$ & $\frac{1}{6} ({\psi_{AB,}}^{\dot A \dot B} R_{CD,\dot A \dot B} - e_{C \dot C}^{\alpha} e_{D \dot D}^{\beta} D_{\alpha} D_{\beta} {\psi_{AB,}}^{\dot C \dot D})$
			\tabularnewline
			& & $+~5~\text{permutations of}~A,B,C,D$
			\tabularnewline
			\hline
			$\phantom{\int^{\int^\int}} \chi_{ABCD} \phantom{\int_{\int}}$ & $-1$ & $-i W_{ABCD}$
			\tabularnewline
			\hline
			$\phantom{\int^{\int^\int}} \lambda_{A \dot A} \phantom{\int_{\int}}$ & $-1$ & $-i C^{\alpha} D_{\alpha} B_{A \dot A} + \text{``}(\psi \psi + e D C) B\text{''} - \frac{i}{4} B_{A \dot A} e^{\beta}_{X \dot X} D_{\beta} C^{X \dot X}$
			\tabularnewline
			\hline
			$\phantom{\int^{\int^\int}} B_{A \dot A} \phantom{\int_{\int}}$ & $-2$ & $\lambda_{A \dot A}$
			\tabularnewline
			\hline
		\end{tabular}
		\par\end{centering}
	\caption{\label{tab:fields} Field content of Witten's topological gravity \cite{Witten:1988xi}.}
\end{table}

In addition to the metric (tetrad), Witten's topological gravity has bosonic fields $C_{A\dot{A}}$ and $B_{A\dot{A}}$ and fermionic fields $\lambda_{A \dot{A}}$, $\psi_{AB\dot{A}\dot{B}}$ and $\chi_{ABCD}$, where the dotted and undotted indices are the $SU(2)_L \times SU(2)_R$ spinor indices in four dimensions. The transformations of these fields, summarized in Table~\ref{tab:fields}, determine the conditions the bosonic backgrounds must satisfy in order to have a BRST-invariant vacuum. As in supersymmetry, these conditions are obtained by requiring that the variations of the fermionic fields vanish.
Included in the variations is the condition $\delta \chi_{ABCD} = W_{ABCD} = 0$, which implies that the universe in phase I must be conformally half-flat. This will be discussed further below.

Using this variation of the fermionic field $\chi_{ABCD}$, one can also see how to write the Lagrangian of this topological gravity theory in the form \eqref{WWaction}:
$$
\{ Q , \chi_{ABCD} W^{ABCD} + \ldots \} = W_{ABCD} W^{ABCD} + \ldots \,,
$$
Note, that action of $Q$ increases the ghost number by $+1$. Therefore, if one wants to construct a manifestly BRST invariant Lagrangian $\mathcal{L} = \{ Q, \mathcal{O} \}$ of ghost number 0, then $\mathcal{O}$ must have ghost number $-1$. In particular, it must involve fields $\chi_{ABCD}$, $\lambda_{A \dot A}$ and $B_{A \dot A}$, which are the only fields of negative ghost number. A quick glance at Table~\ref{tab:fields}, though, makes it clear that the total number of fields $\lambda_{A \dot A}$ and $B_{A \dot A}$ is preserved under the action of $Q$. Therefore, if one wants to construct a purely gravitational term, such as the Einstein-Hilbert term $\mathcal{R}$, then $\mathcal{O}$ can not involve $\lambda_{A \dot A}$ or $B_{A \dot A}$.
Also, $\mathcal{O}$ can not involve $C_{A\dot A}$ and $\psi_{AB, \dot A \dot B}$ because the total number of these fields can only be increased --- but, importantly, never decreased --- by the action of $Q$. This proves that $\mathcal{O}$ must be of the form $\chi_{ABCD} f^{ABCD}$, where $f^{ABCD}$ is a function of the metric and its derivatives (including $W^{ABCD}$), and all such options can be easily classified. In particular, by going through the list of possible functions $f^{ABCD}$ it is easy to see that the Einstein-Hilbert term $\mathcal{R}$ does not appear among the manifestly $Q$-exact terms.

This still leaves a possibility that the Einstein-Hilbert term $\mathcal{R}$ can be written as part of the $Q$-closed but not $Q$-exact action, {\it i.e.} appear in $Q$-cohomology. However, these options too can be easily classified and ruled out. In particular, the BRST variation of the Einstein-Hilbert term is given by:
\begin{equation}
    [Q,\sqrt{g} \mathcal{R}] = 2i \sqrt{g} e^{\alpha A \dot{A}}e^{\beta B \dot{B}}(D_\alpha D_\beta \psi_{AB\dot{A}\dot{B}} - R_{\alpha\beta}\psi_{AB\dot{A}\dot{B}}).
    \label{EHvariation}
\end{equation}
Let us focus on the first term which can be canceled if we find $V$ such that $[Q,V] \propto e^{\alpha A \dot{A}}e^{\beta B \dot{B}}D_\alpha D_\beta \psi_{AB\dot{A}\dot{B}}$. Since a suitable candidate for $V$ should have a variation quadratic in derivatives and linear in the field $\psi$, the same arguments as above tell us that $V$ must be $i)$ either entirely made of the vierbein, or $ii)$ be of the form $\chi \psi \ldots$ where ellipses involve only the vierbein and (possibly) its derivatives. The first option is ruled out because the variation of the vierbein does not produce new covariant derivatives, two of which have to be present in the variation of $V$, and $De = 0$. Option $ii)$ is also ruled out because the variation of $\chi_{ABCD}$ produces $W_{ABCD}$ which already contains two derivatives. These derivatives, however, can not be moved to act on $\psi$, which we need in order to cancel the first term in \eqref{EHvariation}. Moreover, with no extra derivatives in the ellipses, the index structure of such $\chi \psi \ldots$ term does not allow for a Lorentz invariant combination. Therefore, we conclude that the Einstein-Hilbert term cannot be added to the topological action and, in particular, is not generated by quantum corrections. The absence of Eintein-Hilbert term in topological gravity is thus rather different from the current phase of the universe where this term dominates.

As already noted, in addition to the fields discussed above, in order to give the fields a conventional kinetic term it was proposed in ~\cite{Witten:1988xi} that topological gravity is coupled to topological matter and a topological invariant field $\Phi$ couples to some of the fields in the topological theory whose vev
$\langle \Phi \rangle =v_0^2$ will give rise to the desired kinetic term.  This term breaks scale invariance, but we will be assuming that the vev $v_0$ is sufficiently small, as not to invalidate the scaling symmetry of the topological theory.

Gravitational theories of such a topological nature have an intriguing physical interpretation (see, for example, Section 6 in \cite{Witten:1988ze}) and are believed to be confined phases of gravity where general covariance is unbroken. Once the metric acquires an expectation value (i.e. there is a background spacetime) then this symmetry is spontaneously broken and local gravitational excitations, gravitons, emerge. Here an analogy can be made with gauge theories where the confined phase (of QCD for example) has an unbroken local symmetry and does not contain massless gauge bosons.

Finally, there is the question of observables in topological gravity. As in all topological theories, these would be position independent expectation values of operators in $Q$-cohomology. In addition, the absence of spin-2 excitations implies the absence of tensor modes in cosmological observables.

\subsection{Flatness}

We are assuming no observables of frame I will distinguish positions, so the metric should be homogeneous, i.e. a constant curvature metric.
We note that the time direction is picked out as an invariant concept in both phases.  We would like to determine the consequences of this for the geometry in phase I as viewed from the frame II perspective. Therefore, the most general metric with these symmetries is
\begin{equation}
ds^2=-dt^2+a^2(t)\left[\frac{dr^2}{ (1-k r^2)} +  r^2 d\Omega^2\right]
\label{FRWansatz}
\end{equation}
where $k=+1,0,-1$ for positive, flat or negative curvature spaces.
However, as discussed above, the solutions to BRST invariant configurations in 4d topological gravity are conformally flat self-dual geometries, which have
\begin{align}
W_{ABCD}=0.
\end{align}
This condition by itself allows all three possibilities above. We will view time as a continuous element between phase I and phase II, whereas spatial geometry will in general undergo duality. Thus, a natural assumption is that the metric can be expressed as a {\it flat metric up to a conformal factor that is only dependent on time}, which is the only duality invariant coordinate. This is of course equivalent to having an FLRW metric \eqref{FRWansatz} with $k=0$, {\it i.e.}
\begin{align}
    ds^2 = a^2(\eta)(-d\eta^2+ dx^i dx^i)
\end{align}
Moreover in phase II, since the metric should smoothly connect, we learn that at the beginning of the FLRW cosmology, the universe is spatially flat.
It would be interesting to further study the interplay between the BRST invariance of the topological phase I and the assumption of duality we have imposed on the metric, leading to a solution of the flatness problem.

\subsection{Scale invariance}

The gravity sector of topological gravity is scale invariant, as already discussed.   In particular the Einstein-Hilbert term $\int \sqrt{g} R$ is absent from the action due to BRST invariance.   However, this is not the full story.  First of all the coupling constant of the topological gravity action, the term in front of $W^2+\ldots$ generally runs with scale and is not scale invariant.
Consider for example a nearly conformal field theory coupled to gravity.  We know that in a gravitational background, this leads to violation of scale invariance due to the conformal anomaly which causes running of the $W^2$ term through the $c$ coefficient.
The conformal anomaly is captured by the $a$ and $c$ coefficients (see~\cite{Duff:1993wm} and references therein).  In particular we have
\begin{align}
{T}{^\mu_\mu} = \frac{1}{16\pi^2}\left(
c W^2 - a E \right)
\end{align}
where $E$ is the Euler density.
If we denote the dilatation mode of the scale invariant theory by a field\footnote{This field $\tau$ will be identified with the gauge-invariant scalar perturbation $\zeta$ in the FLRW phase.} $\tau$, where the metric can be viewed as a fixed metric $\hat g$ scaled by $g=e^{2\tau} \hat g$, we have a running of the coupling in the gravitational action given by the coupling of the dilaton field $\tau$ to the $W^2$ term and the rest of the topological action.

Notice that there was an overall scaling ambiguity in Witten's topological gravity Lagrangian, which can be viewed as fixing the coupling constant of the theory $1/g^2$. In addition, we can also add the Euler density which is topological with a strength $\kappa$. We thus end up with the following contributions to the gravitational part of the action\footnote{We are in Euclidean space with the path integral weighted by $e^{-S}$.}
\begin{align}
\label{eq:W2action}
    S &= S_{top}+ S_{\rm anomaly} + \ldots\\
\label{eq:topaction}
    S_{top} &= \int \left[\frac{1}{g^2} W^2 + \kappa E_4 +...\right]\\
\label{eq:KSaction}
   S_{\rm anomaly} &= \frac{1}{16\pi^2}\int
   \Bigg[
   - \tau
   \left(c  W^2 - a E_4 \right)  \\ & \nonumber \quad \quad
   +a \left(4(R^{\mu\nu} -\frac{1}{2}g^{\mu\nu} R)\partial_\mu \tau\partial_\nu \tau
    -4(\partial \tau)^2 {\mathop{}\!\mathbin\Box} \tau
    +2(\partial \tau)^4
    \right)
        \Bigg]
\end{align}
The action $S_{\rm anomaly}$ was derived some time ago in~\cite{Fradkin:1983tg} and was more recently used by Komargodski and Schwimmer ~\cite{Komargodski:2011vj} in the proof of the $a$-theorem.
In a sense the field $\tau$ can be viewed as a BRST invariant dynamical coupling constant of the theory which the topological theory allows.  For example it can be viewed as being related to the topological invariant field $|\Phi |=v_0^2\ {\rm exp} (-2\tau)$ in Witten's theory, and integrating this field out will generate the effective terms in $S_{\rm anomaly}$.   This in particular guarantees that $S_{\rm anomaly}$ can be completed to a BRST invariant part of the topological theory.
Note that the $1/g^2$ can be identified with the expectation value of $-c\tau/16\pi^2$ so that the above can be considered an action for `dilaton-conformal gravity' with $\tau$ playing the role of the dilaton.    Note that this identification, due to the fact that $c>0$ which is a consequence of unitarity, implies that at larger distances where $\tau$ increases, it corresponds to decreasing $1/g^2$, or making $g$ bigger.  In other words the coupling constant of this theory gets stronger in the IR.

We will be interested in the two-point function of the scalar fluctuations, $\tau(x)$. Formally the fact that this field is BRST invariant suggest that the correlation functions of it in the topological theory are position independent.  In agreement with this expectation, note that classically $\tau$ is a dimensionless field in equation~\eqref{eq:W2action} above and thus, its two-point function is classically scale invariant. This would naively suggest as we expect a position independent correlation.  However, due to conformal anomaly, this is not the case, which presumably should also lead to a mild breaking of topological invariance.  We can estimate the two point function of $\tau$ coming from conformal anomaly as follows.
The quantum correction to the two-point function arises due to the non-zero trace  of the stress energy tensor,
\begin{align}
    x^\mu \frac{\partial}{\partial x^\mu}
    \llangle \tau(x) \tau(0) \rrangle
    &= \llangle \left(\int d^4 y \, T^\mu_\mu \right)
    \tau(x) \tau(0) \rrangle
    =
\frac{1}{16\pi^2}
\llangle
\int (c W^2 - a E) \tau (x) \tau (0) \rrangle
    \,.
\end{align}
We approximate the 3-point function above by its disconnected piece
\begin{align}
   \llangle \int (c W^2 - a E) \tau (x) \tau (0) \rrangle
\simeq
\llangle \int (c W^2 - a E) \rrangle
\llangle \tau (x) \tau (0) \rrangle
\end{align}
At weak coupling, which as we will argue our cosmology corresponds to, we expect that
\begin{align}
    \langle W^2 &\rangle \sim g^2
    \\
    \langle E \rangle &\sim \exp(-1/g^2)
    \,.
\end{align}
Note that the vev of the topological term $E$, which classically is $\langle E \rangle \sim 0$ around our flat space ansatz, receives contributions only from non-trivial topologies and is thus highly suppressed at weak coupling.  This is familiar in the context of gauge theories with the role that $E$ plays here for us being played by the topological term ${\rm tr} F\wedge F$. This leads to
\begin{align}
x^\mu \frac{\partial}{\partial x^\mu}
    \llangle \tau(x) \tau(0) \rrangle
    \simeq
cg^2
\llangle \tau (x) \tau (0) \rrangle
\end{align}
where we have ignored the contribution of $E$ as it is negligible in the weak coupling limit.
To cast the two-point function in a more familiar form, we can take the spatial Fourier transform of $\langle \tau(x) \tau(0) \rangle$, evaluated at equal times,
\begin{align}
\llangle \tau (0,\vec{x}) \tau(0,0)\rrangle
\equiv
\int \frac{d^3 k}{(2\pi)^3}
\llangle \tau (\vec{k}) \tau(-\vec{k}) \rrangle'
\exp(-i\vec{k}\cdot \vec{x})
\end{align}
where we have defined $\langle\ldots \rangle'$ as the correlator with the momentum conserving delta-function stripped off.  We will drop the $'$ in the following, keeping in mind that that is what we mean by the correlator.
The 3D-Fourier transformed version of the dilatation equation is,
\begin{align}
    -\left(3+ \vec{k}\cdot\frac{\partial}{\partial \vec{k}}\right)
    \llangle \tau(\vec{k}) \tau(-\vec{k}) \rrangle
    &= \frac{c g^2}{16\pi^2}
    \llangle \tau(\vec{k}) \tau(-\vec{k}) \rrangle
\end{align}
which leads to the following
familiar form for the scalar fluctuation power spectrum,
\begin{align}
\Delta^2(k)
\propto
    \frac{k^3}{2\pi^2}
    \llangle \tau(\vec{k}) \tau(-\vec{k}) \rrangle
    &\propto
    \left(\frac{k}{k_*}\right)^{n_s-1}
\end{align}
with $n_s =1-\frac{cg^2}{16\pi^2}$, and $k_*$ is the reference pivot scale. Note that the red-tilt of the two-point function here is a direct consequence of $c>0$, which is a consequence of the unitarity of the theory.

The amplitude for $n$-point functions for the field $\tau$ can also be estimated using the Lagrangian for the $\tau$ field in equation~\eqref{eq:W2action}.
However, there is no canonical kinetic term for $\tau$ in the anomaly functional $S_{\rm anomaly}$.  One way to estimate the magnitude of the two point function is to note that in Witten's theory $\tau$ may be related with the expectation value\footnote{ In order for this vev not to affect the predictions of the nearly scale invariant theory we are considering we need to assume $v_0 \ll k$ for scales $k$ of interest.} of a BRST invariant operator $\Phi \sim {\rm exp} (-2\tau)$.  If we consider $\Phi$ as a massless free field, then this leads to an on-shell relation of the form $\square \tau =(\partial \tau)^2$ (as in \cite{Komargodski:2011vj}) and thus the quartic term in $S_{\rm anomaly}$ is compatible with a term\footnote{Topological gravity is unitary as noted in  \cite{Witten:1988xi} despite the fact that conformal gravity by itself is not unitary.  It is natural to expect that a term such as $(\square \tau) ^2$ which in usual theories are not unitary, in the context of topological theory could be part of a unitary theory.} $a(\square \tau) ^2$.
In particular, we see that $a$ sets the amplitude of $\tau$ correlators which lets us estimate the scalar power spectrum amplitude, so that,
\begin{align}
    \Delta^2(k)
    = \frac{k^3}{2\pi^2} \langle \tau(\vec{k})\tau(-\vec{k}) \rangle
    \simeq \frac{1}{a} \left(\frac{k}{k_0}\right)^{-cg^2/16\pi^2}
\end{align}
Therefore, the small value of the power-spectrum is explained by the large central charges $a \sim c \sim 10^{10}$ (see~\cite{Hofman:2008ar} for arguments implying $a\sim c$). At the same time, the near-scale invariance is explained by the weak coupling, so that $cg^2/16\pi^2 \simeq 0.03$ and thus $g\sim 10^{-5}$.
 Thus, the theory is at weak coupling and has a large central charge.\footnote{The fact that the theory has a large central charge resonates with the speculations that in a topological theory of string, all the different string modes will be viewed as degenerate massless fields.}

We can also consider higher point fluctuations. The universal nature of the Weyl anomaly requires the presence of the interaction terms shown in~\eqref{eq:W2action} which contribute to higher point correlation functions.
We point out that the amplitude of $f_{\rm NL}$ is not suppressed by $g$ or $a^{-1} \sim c^{-1}$. Focusing on the scaling of the 3-point function with these parameters we have:
\begin{align}
\langle \tau(\vec{k}_1)\tau(\vec{k}_2)\tau (\vec{k}_3)\rangle
&\sim
a \prod_{i=1}^3 \langle \tau(\vec{k}_i) \tau (-\vec{k}_i) \rangle
\end{align}
The prediction for non-Gaussianity parameterized by $f_{\rm NL}$ is then
\begin{align}
    f_{\rm NL}
    &=
    \frac{\langle \tau(\vec{k}_1)\tau(\vec{k}_2)\tau (-\vec{k}_1-\vec{k}_2)\rangle}
    {\langle \tau(\vec{k}_1)\tau(-\vec{k}_1)\rangle
    \langle \tau(\vec{k}_2)\tau(-\vec{k}_2)\rangle
    }
    \sim \mathcal{O}(1).
\end{align}
That $f_{\rm NL} \sim \mathcal{O}(1)$ is a consequence of the fact that the same parameter $a$ sets the size of the two- and higher-point functions. This is in contrast to inflation where the simplest models have $f_{\rm NL} \sim \mathcal{O}(\epsilon)$ as we discuss in Section~\ref{sec:compar} below. Using the in-in formalism, one can calculate the magnitude and shape of the expected non-Gaussianities. Since the three-point interaction is proportional to $\Box \tau$, it accidentally vanishes on-shell and so will its contribution to the three-point function. However, for the topological phase to end, our background must have some time evolution which could give $\tau$ particles slightly off-shell momenta. This could lead to a small contribution to the three-point function. Since a precise determination of the amplitude of $f_{\rm NL}$ would depend on model details, we leave the investigation of this and other potential effects for future work.

The four-point function given by the above interaction does not vanish on-shell and provides a universal prediction. The contributions of the various terms in $(\pd \tau)^4$ to the four-point function is given by the sum of the three terms:
\begin{align*}
  T^{\dot{\tau}^4} &= \frac{1}{a^3}\frac{96}{(k_1 k_2 k_3 k_4)^2}\frac{1}{k_1+k_2+k_3+k_4}\\
  T^{\dot{\tau}^2(\pd_i\tau)^2} &= \frac{1}{a^3}\frac{-32}{(k_1k_2k_3k_4)^3} \frac{k_1 k_2 (\mathbf{k_3\cdot k_4}) + 5\;\mathrm{perm.}}{k_1+k_2+k_3+k_4}\\
  T^{(\pd_i \tau)^2(\pd_j\tau)^2} &= \frac{1}{a^3}\frac{16 }{(k_1k_2k_3k_4)^3} \frac{(\mathbf{k_1\cdot k_2}) (\mathbf{k_3\cdot k_4}) + 5\;\mathrm{perm.}}{k_1+k_2+k_3+k_4}
\end{align*}
The $a$-dependence again gives a prediction\footnote{Here $\tau_{\rm NL}$ is a parameter measuring a contribution to the 4-pt function and is not to be confused with the field $\tau$.} of $g_{\rm NL} \sim \tau_{\rm NL} \sim 1$. While this may be difficult to detect with near-future experiments, forecasts~\cite{Bartolo:2015fqz} have shown that a cosmic variance experiment measuring correlations between CMB temperature anisotropies and spectral distortions can in principle detect $g_{\rm NL} \sim 0.4$.

\section{Contrasting topological early phase with Inflation}
\label{sec:compar}
\begin{figure}
\centering
\includegraphics[width=\textwidth]{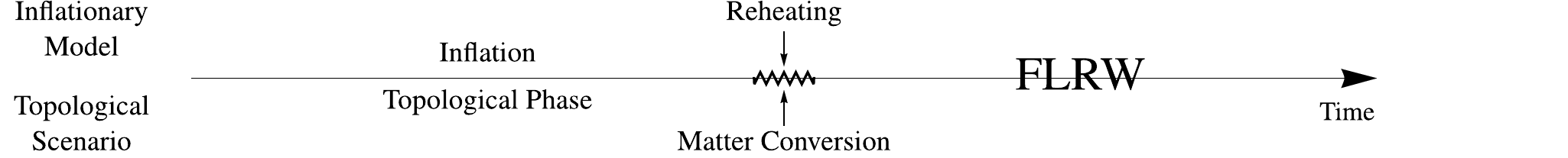}
\caption{Comparison between the inflationary and topological paradigms for the early universe. The topological scenario replaces the period of accelerated expansion by a topological phase to explain homogeneity, isotropy, flatness and near scale-invariance. The degrees of freedom of our universe are repopulated by a period of reheating after inflation. In the topological scenario, matter composed of degrees of freedom that are light in a dual description is converted to our degrees of freedom and populates our universe. In both cases, the universe for $t>0$ is well described by a Hot Big Bang cosmology.}
 \label{fig:comparison}
\end{figure}

In this section we compare and contrast our scenario with the inflationary scenario.  On the surface of it there are a number of common features between the two approaches  as can be seen in Fig.~\ref{fig:comparison}.

Of course, the end result for both is the FLRW scenario.  Both of them involve a kind of phase transition:  in the case of inflation the transition is marked by the end of the inflation and reheating as the inflaton settles to the minimum of the potential.  In the case of the topological scenario the phase transition is marked by the conversion of the matter from phase I to phase II.   In both scenarios this leads to a nearly homogeneous thermal initial condition for FLRW.   In both scenarios the homogeneity of the space is described by a novel phenomenon:  in the inflationary scenario by the exponential expansion of the space and in the topological phase by the fact that gravity is described by a topological theory.  In the inflationary scenario the fluctuations of the inflaton field leads to scalar fluctuation, whereas in the topological phase which involves only global/zero modes and only through scale anomalies do we get fluctuations in the otherwise thermal background.  Below we first give a theoretical analogy to QCD-like theories and how these two approaches compare in that context.  We then proceed to highlight some of the distinct predictions of these two approaches.

\subsection{An analogy with a QCD-like theory}
We draw a very rough analogy with QCD to highlight the qualitative difference between the topological scenario and inflation.
In the FLRW universe, as we go to earlier times, we get to higher energy densities, and stronger gravitational interactions. The analogy of going to earlier times is RG flow towards the IR in a QCD-like theory.

In QCD, the coupling runs to strong values in the IR, and to describe the theory below $\sim 1$ GeV scale we have to resort to the dual chiral Lagrangian description. The chiral Lagrangian is nonetheless predictive since chiral symmetry is weakly broken, and dictates the physics of processes involving the mesons which are pseudo-Nambu Goldstone bosons of chiral symmetry.

There is an alternative possibility. Before we hit strong coupling, the QCD theory could have been Higgsed down to the Coulomb phase (at say the 10 GeV scale). Then we do not ever hit a strongly coupled regime, and perturbation theory is reliable in the IR. We also have a Higgs field that breaks the chiral symmetry, giving rise to the usual octet of pion fields. As in QCD, the chiral symmetry breaking pattern sets most of the pion physics.

Inflation is analogous to the linear sigma model, where before we hit the gravitational strong coupling scale we change into a different weakly coupled phase, and calculations for early and late universe can be made in one single duality frame.\footnote{Indeed to make the analogy even more complete note that it has been argued that Einstein-Hilbert gravity around some fixed background can be interpreted as a spontaneously broken phase of gravity, with the massless graviton thought of as a kind of a Nambu-Goldstone boson~\cite{Witten:1988ze}.}

On the other hand, the topological early universe is analogous to real world QCD, where as we go to the early universe we run into strongly coupled gravity.
In a phase with unbroken general covariance, we expect there to be no local physics, and the theory is expected to be topological.
From this point of view, our ansatz of the strongly coupled early phase of the universe as a topological gravity theory is natural.
Our calculations of the scalar fluctuations above follow from the symmetries of such a phase of gravity. As mentioned above, we do not get any tensor mode fluctuations in our topological phase. This is entirely analogous to the fact that we have no massless spin-1 gluons in the confined phase of QCD. In inflation we of course do have tensor modes, reflecting that we stay in a weakly coupled phase as in the Coulomb phase of the linear sigma model case above.

From this point of view, the relative phenomenological success of inflationary models is explained by the approximate scale symmetry of the early topological phase. This also explains one of the puzzles of inflationary models -- why are there no tensor modes detected so far? It may be that the tensor modes are just around the corner, or indeed beyond our experimental reach (although this would typically involve tuned models of inflation).   However, if our model is a valid description of the universe, topological early phase explains why there are no tensor modes fluctuations, and if we were to detect any it would falsify this possibility.

\subsection{Predictions of inflation and the topological phase}

We briefly recall here the standard phenomenological predictions of  inflation and contrast them with the predictions of the topological phase.
There are a number of excellent reviews on inflation and models that realize it (see e.g.~\cite{Baumann:2009ds} and references therein). The current constraints on various models from {\it Planck} can be found in~\cite{Akrami:2018odb}. The phenomenology of inflationary models is vast, but it can be captured model-independently in the language of EFT of inflation~\cite{Cheung:2007st}. This underlines the fact that phenomenological predictions of inflation are driven by the symmetry breaking pattern of spacetime isometries in the early universe.
The universal part of the Goldstone action in the inflationary EFT takes the form:
\begin{align}
\label{eq:eftofinflation}
    \mathcal{L} = \Mpl^2 H^2 \epsilon \left(\dot{\pi}^2 - \frac{(\partial_i\pi)^2}{a^2}\right)
\end{align}
in the ``Goldstone boson equivalence limit" (see~\cite{Cheung:2007st} for details). This determines the amplitude of the two-point function. The tilt of the two-point function follows from the slow variation of $H(t)$
in the quasi-de Sitter background. As we show above, the topological phase also predicts a scale-invariant two-point function. In our case, the tilt is due to the Weyl anomaly of the theory and the condition $c>0$ predicts the red tilt of the spectrum.

Non-Gaussianities in the power spectrum arise from interaction terms in the EFT Lagrangian. In the simplest slow-roll inflation, the gravitational interactions produce a non-Gaussianity which is suppressed by the slow-roll parameter,
\begin{align}
    f_{\rm NL} \sim \mathcal{O}(\epsilon)
    \,.
\end{align}
In theories with large higher-derivative interactions, it is possible to get larger values of $f_{\rm NL} \sim 1$. However, these theories need a UV completion not far from the Hubble scale. As discussed above, time evolution in the topological phase could induce a non-zero $f_{\rm NL}$ that would not be suppressed by a `slow-roll'-type parameter. However, in the simple case considered in this paper, $f_{\rm NL}$ accidentally vanishes.

The tensor power spectrum in inflationary models only depends on the Hubble scale during inflation
\begin{align}
    \Delta_\gamma \propto \frac{H^2}{\Mpl^2}
\end{align}
and is a universal prediction of the inflationary paradigm, directly tied to weakly coupled gravity during inflation.
This is in stark contrast with the model presented in this paper, where the topological nature of the early phase predicts no tensor modes.

\section{Future directions}
\label{sec:future}
The topological scenario we have proposed can have many possible realizations.  In this paper we have focused on one specific example of it, given by Witten's topological gravity.  It is likely that there are other equally interesting models of topological gravity that could be interesting for applications in early universe cosmology.  Nevertheless the fact that in the context of Witten's topological gravity theory, we have found a link between conformal anomaly, unitarity, and the IR tilt of the power spectrum is quite reassuring.

There are a number of aspects of the topological theory we have considered that would be interesting to develop further.  In particular the connection between conformal anomaly and breaking of topological invariance in phase II is natural to expect in this model.  This should be further developed.  Also we should develop techniques to compute correlation functions of the scaling field $\tau$ more precisely, which we have only estimated in this paper.

Finally, even though we have emphasized string theory dualities as a main motivation for considering a topological phase for the early universe, it would be important to describe this phase in more intrinsically stringy terms, and not just in terms of an effective action.  We leave developing such connections to future work.

\subsubsection*{Acknowledgements}

We have greatly benefited from discussions with Z. Komargodski and A. Schwimmer.

PA was supported in part by NSF grants PHY-1620806 and PHY-1915071, the Chau Foundation HS Chau postdoc support award, the Kavli Foundation grant Kavli Dream Team, and the Moore Foundation Award 8342. The work of S.G. is supported by the U.S. Department of Energy, Office of Science, Office of High Energy Physics, under Award No. DE-SC0011632, and by the National Science Foundation under Grant No. NSF DMS 1664227 and the work of G.O. and C.V. is supported in part by a grant from the Simons Foundation (602883, CV).

\bibliographystyle{utphys}
\bibliography{topologicalCosmology}

\end{document}